# NUMERICAL AND EXPERIMENTAL INVESTIGATION OF A THREE-AXIS FREE ROTATION WIND TUNNEL MODEL

Muller L.[(1)], Libsig M.[(2)], Martinez B.[(3)], Bidino D., Bastide M., Baily Y.[(4)], Roy J.-C.[(5)]

[(1)] ISL, 5 rue du Général Cassagnou, 68300 Saint-Louis (France), laurene.muller@isl.eu
[(2)] ISL, 5 rue du Général Cassagnou, 68300 Saint-Louis (France), michel.libsig@isl.eu
[(3)] ISL, 5 rue du Général Cassagnou, 68300 Saint-Louis (France), bastien.martinez@isl.eu
[(4)] FEMTO-ST, 2 Avenue Jean Moulin, 90000 Belfort (France), yannick.bailly@univ-fcomte.fr
[(5)] FEMTO-ST, 2 Avenue Jean Moulin, 90000 Belfort, France, jean-claude.roy@univ-fcomte.fr

**ABSTRACT**

The current need of improving performance in terms of control and aerodynamic efficiency of ammunitions leads to the necessity of performing accurate flying geometry characterizations. Therefore, new investigation methods are developed in order to increase the aerodynamic knowledge. Free flight measurements experiments are the most common way to obtain dynamic aerodynamic coefficients. However, they do not always allow neither easy nor perfect measurements conditions. Currently ISL develops a stereovision method based wind-tunnel measurements methodology for investigation of a 3-axis free rotation model. This methods has been applied to the DREV-ISL reference model [1] [2] [3] [4] [5] in order to compare coefficients obtained by this method with numerical results.

## 1. INTRODUCTION

For concept validation and pitch damping coefficient measurements in wind tunnel, a three-axis free rotating test bench for projectiles, called MiRo [11], is under development at the ISL (French-German Research Institute of Saint-Louis). The final goal of this setup would be to become able to investigate the attitude of spin-stabilized models fitted with uncoupled actuators. Due to the mechanical complexity of such a device, the development is performed step by step. This paper presents the first step for which a methodology to obtain static and dynamic aerodynamic coefficients on a free-rotating finned model has been developed. The measurement of the motion of the body during the wind tunnel test is performed with a stereovision technique, for which two high-speed cameras are employed simultaneously. Afterwards, for a stereoscopic purpose, both recordings are processed frame by frame and coupled by means of an image processing technique. At the end of the process, the attitude of the projectile during the wind tunnel test is reproduced numerically in order to identify the pitch damping moment coefficient with a curve fitting algorithm based on a theoretical motion model. To increase the confidence level of the obtained measurement, which can especially be affected by a cavity effect that is directly linked to the test bench principle, drag, lift and pitch moment terms have been compared for different configurations. Due to a very limited space inside the test bench and no optical access, the cavity effect investigation has exclusively been processed by RANS CFD simulations up to a 2° angle of attack. Previous wind tunnel and free flight results from literature have been taken as a comparative basis for the validation of both CFD results and MiRo wind tunnel measurements.

## 2. DESCRIPTION OF THE EXPERIMENTAL SETUP

### 2.1. THE MECHANICAL PRINCIPLE OF THE MIRO TEST BENCH

The MiRo test bench [17] consists in holding the model from the rear at its center of gravity while allowing the rotations to be free in the 3 directions in space. The roll motion is obtained by means of two bearings situated on both sides of a Cardan-like joint, which allows both pitch and yaw motions (See Fig. 1).

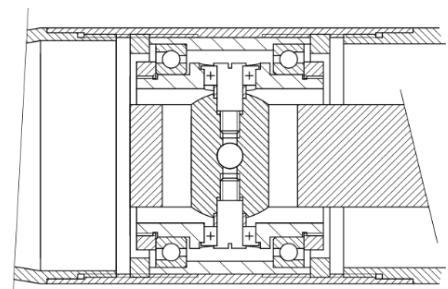

Figure 1. Motion device

A position-locking device, which is implemented around the Cardan-like joint, allows to keep the model in a secure position as long as the blow down is not established. A cavity has to be created at the rear of the model for a sting to flush mount the inner part of the Cardan joint on the wind tunnel's structure, so that the projectile can be maintained in the flow while beeing free in all 3 directions of rotation. For this reason, the pitch and yaw amplitude of the model is limited to 2 degrees. However, this moderate amplitude is clearly enough for projectiles to be characterized on a large part of their flight trajectory. Due to the mechanical complexity of the device, the development is performed step by step. In order to avoid instability problems, and due to a large amount of already existing data, the finned DREV-ISL rocket [1] [2] [3] [4] [5] which design is shown in Fig 2 was retained. For mechanical and ISL trisonic wind tunnel constraints reasons, the calibre of the test model is set to 40 mm, both for the experimental and the numerical investigations.

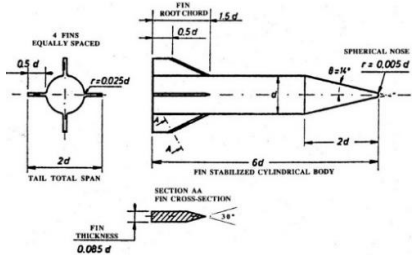

Figure 2. DREV-ISL model

For this test, the centre of gravity was located at 180.7 mm with respect to the base and its inertia was $9.3 \cdot 10^{-4}$ kg.m².

## 2.2. THE ATTITUDE DETERMINATION

### 2.2.1. THE STEREOVISION PRINCIPLE

In order to record the motion in all directions of space, the experimental setup has to be able to capture the depth of the scene, like it is done by the human brain by combining information from both eyes. A one-eyed person, for whom the configuration is similar to a unique camera, is not able to see in 3D. For the stereovision technique that is employed herein, the same principle is recreated computationally. Therefore, the computer needs at least two cameras in order to obtain two different views of the scene. So as to follow the motion and reconstruct the flight, markers are placed on the model in order to be recognized by means of an image treatment process. Therefore, both image series are processed and coupled frame by frame. At the end of the process, the attitude of the projectile during the wind tunnel test is reproduced numerically in order to obtain the pitching moment derivative coefficient $C_{m\alpha}$ and the pitch damping moment coefficient $C_{mq}$.

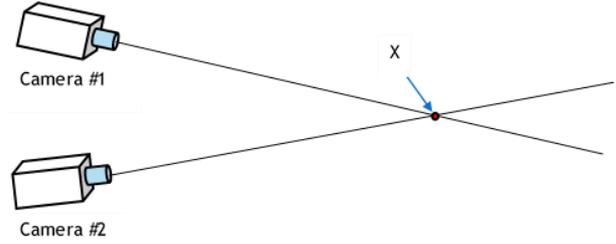

Figure 3. Stereovision principle

### 2.2.2. MATHEMATICAL STEREOVISION MODEL

The mathematical model which is employed for the stereoscopic determination of the attitude of the wind tunnel tested device is based on the pinhole camera model. The pinhole camera is a black box, that contains an aperture like a small hole and which reproduces an image after the passage of the light through the orifice. The mathematical model (Eq. 5) describes the relationship between the 3D coordinates of point M in space and its projection onto an image point m of an ideal pinhole camera as shown on the computer screen and for which the coordinates are expressed in pixels (See Fig. 4) [6]. The model does not consider the geometric distortion.

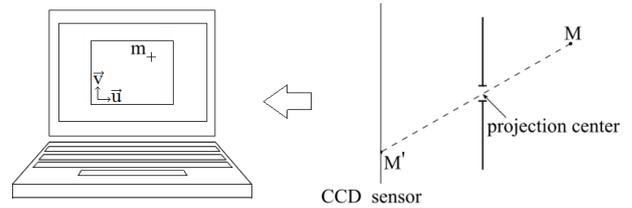

Figure 4. Pinhole model representation

The relationship could be decomposed into three successive elementary transformations as shown in Fig. 5. For each reference frame $(\vec{A}, \vec{B}, \vec{C})$, the notation of the associated coordinates is A, B and C, respectively.

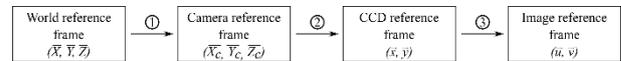

Figure 5. Pinhole model decomposition

The transformation ① (See Eq.1) transforms the physical 3D point M (object points) expressed in the world reference frame $(\vec{X}, \vec{Y}, \vec{Z})$ to the camera reference frame $(\vec{X_c}, \vec{Y_c}, \vec{Z_c})$ thanks to a rotation matrix $R_{3\times3}$ ($r_{i,j}$), whose elements are expressed with the Euler angles, and a translation vector $T$ ($t_x$ $t_y$ $t_z$).

$$\begin{pmatrix} X_c \\ Y_c \\ Z_c \\ 1 \end{pmatrix} = \begin{pmatrix} R & T \\ 0_{1\times3} & 1 \end{pmatrix} \begin{pmatrix} X \\ Y \\ Z \\ 1 \end{pmatrix} \quad (1)$$

The transformation ② (See Eq.2) projects the object point M onto the CCD plane $(\vec{x},\vec{y})$ by perspective projection. This operation gives the projected point M'. In this equation, $f$ is the focal length and $s$ is a scale factor depending among other on the distance between the pinhole and the object point M.

$$s\begin{pmatrix}x\\y\\1\end{pmatrix}=\begin{pmatrix}f&0&0&0\\0&f&0&0\\0&0&1&0\end{pmatrix}\begin{pmatrix}X_c\\Y_c\\Z_c\\1\end{pmatrix} \quad (2)$$

The transformation ③ (See Eq.3) transforms the projected point M' from the CCD frame of reference $(\vec{x},\vec{y})$ into the image coordinates system $(\vec{u},\vec{v})$. The obtained image point m which coordinates express its position in pixels on the recorded pictures that are displayed on the computer. Thereby, the $p_u$ and $p_v$, horizontal and vertical pixel to length ratio, and $(u_0, v_0)$, the location in the picture of the intersection between the CCD plane and the optical axis passing through the pinhole, are employed as shown in Eq. 3.

$$\begin{pmatrix}u\\v\\1\end{pmatrix}=\begin{pmatrix}p_u & p_u\cot(\theta) & u_0+u_v\cot(\theta)\\0 & p_v/\sin(\theta) & u_v/\sin(\theta)\\0 & 0 & 1\end{pmatrix}\begin{pmatrix}x\\y\\1\end{pmatrix} \quad (3)$$

$\theta$ represents the possible non-orthogonality of the image rows and columns. In this case, we assume that the orthogonality is perfect, hypothesis which is valid with the cameras and lenses that are employed herein, which means $\theta=\pi/2$. Therefore, Eq. 3 becomes:

$$\begin{pmatrix}u\\v\\1\end{pmatrix}=\begin{pmatrix}p_u & 0 & u_0\\0 & p_v & u_v\\0 & 0 & 1\end{pmatrix}\begin{pmatrix}x\\y\\1\end{pmatrix} \quad (4)$$

Finally, by combining Eq.1, Eq.2 and Eq. 4 the mathematical expression of the pinhole model is:

$$s\begin{pmatrix}u\\v\\1\end{pmatrix}=\begin{pmatrix}\alpha_u & 0 & u_0 & 0\\0 & \alpha_v & u_v & 0\\0 & 0 & 1 & 0\end{pmatrix}\begin{pmatrix}R & T\\0_{1\times3} & 1\end{pmatrix}\begin{pmatrix}X\\Y\\Z\\1\end{pmatrix} \quad (5)$$

With

$$\alpha_u=f.p_u \text{ and } \alpha_v=f.p_v \quad (6)$$

The intrinsic parameters ($\alpha_u$, $\alpha_v$, $u_0$ and $v_0$) are specific to the lens of the camera, while the three Euler angles and the three components of the translation vector are the extrinsic parameters that express the camera position with respect to the object. This ten parameters are determined by means of a calibration process described in part 2.2.3.

As the projectile is observed by two cameras and as each single marker creates an image point on both camera recordings, which coordinates are noted $U_1$ and $U_2$, the stereoscopic relationship [17] between each single marker (object point) and its image is described by Eq. 5 and can be written in a more compact way:

$$\begin{cases}s_1U_1=I_{C1}(R_1.X+T_1)\\s_2U_2=I_{C2}(R_2.X+T_1)\end{cases} \quad (7)$$

With

$$U_{1,2}=\begin{pmatrix}u_{1,2}\\v_{1,2}\\1\end{pmatrix}, X=\begin{pmatrix}X\\Y\\Z\end{pmatrix} \quad (8)$$

In this case $s_{1,2}$ is the scale factor related to the camera 1 or 2, $I_{1,2}$ is the intrinsic parameters matrix, $R_{1,2}$ is the rotation matrix and $T_{1,2}$ is the translation vector between the world reference frame and the camera reference frame. All these parameters except for $s_{1,2}$ are determined at the calibration process. Each relation of the system in Eq. 7 is the equation, in matrix form, of a line in 3D space. Thus, the coordinates of the unknown 3D point can be calculated because it represents the intersection of both camera axes lines (See Fig. 3). Eq. 7 is a system of six scalar equations with five unknowns: X (three scalar values), $s_1$ and $s_2$ which leads to an over-determined system of equations:

$$\begin{cases}-I_1.T_1 = I_1.R_1.X-s_1.U_1\\-I_2.T_2 = I_2.R_2.X-s_1.U_2\end{cases}$$

$$\Leftrightarrow \begin{bmatrix}R_1 & -I_1^{-1}.U_1 & 0_{(3)}\\R_2 & 0_{(3)} & -I_2^{-1}.U_2\end{bmatrix}\cdot\begin{bmatrix}X\\s_1\\s_2\end{bmatrix}=-\begin{bmatrix}T_1\\T_2\end{bmatrix} \quad (9)$$

This latter can be rewritten in a compact form (Eq. 10) with $P$, a 6-component vector and $H$, a 6x5 matrix:

$$H.\begin{bmatrix}X\\s_1\\s_2\end{bmatrix}=-P \quad (10)$$

By performing a least square optimization process, it is possible to calculate the 3D coordinates of X as well as $s_1$ and $s_2$. The least square matrix solution can be expressed as:

$$\begin{pmatrix}X\\s_1\\s_2\end{pmatrix}=-(H^T.H)^{-1}.H^T.P \quad (11)$$

### 2.2.3. CALIBRATION PRINCIPLE

A calibration step [15] [16] is required for the initialization of the mathematical pinhole model. This step is divided into two sub-steps: the determination of known 3D object points on the image and an optimization algorithm employed to determine the intrinsic and extrinsic parameters. The first calibration step is performed thanks to an image of a 3D raw card, which is composed of three cube inner faces covered with a checkerboard pattern. The angular positioning of this latter, as well as the size of the squares, are known, so that the ten constant parameters of the pinhole model can be determined. The placement is chosen in such a way that both cameras perfectly see the three faces of the raw card. The algorithm estimates the position of the camera via the checkerboard squares deformation in the image. Three stages are necessary to achieve the calibration:
1. Manual selection of each faces' extreme points, determination of the position of each checkerboard corner by the Harris corner detection algorithm [7], and estimation of the distance between the selected corners.
2. Deduction by direct linear transformation [8] of each control point approximate position on the image and association with its respective 3D coordinates.
3. Determination of the parameters set by the Levenberg-Marquardt optimization method [9], [10].

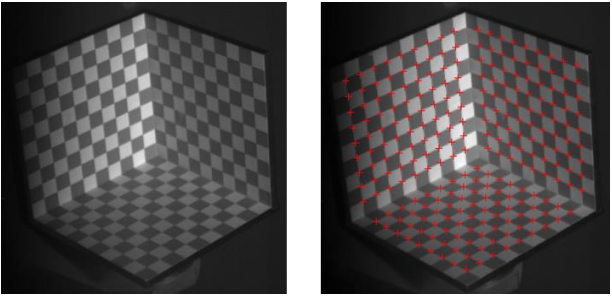

Figure 6. 3D raw card (left) and checkerboard corners detection (right)

During the optimization, the code minimizes the distance between the detected control points on the images and their theoretical image position foreseen by the pinhole model, which depends on the intrinsic and extrinsic parameters.

## 2.3. AERODYNAMIC COEFFICIENT DETERMINATION METHODOLOGY

In order to determine the static and dynamic aerodynamic coefficients, the angle of attack of the model has to be perturbed in order to obtain a damped oscillating attitude, on which the frequency and the amplitude evolution has to be analysed.
The oscillating motion is specific to stable projectiles. If its amplitude is considered as small and the velocity of the flow as constant, the angular motion of the projectile can be described by the linearized equations of incidence as given by McCoy [14]. Additionally, these equations can be simplified thanks to the test conditions:
- No gravity effect
- Spin rate close to zero
- No angular variation of the velocity vector

Under these conditions, the angle of attack evolution as a function of time is a damped oscillating motion defined as:

$$\alpha = e^{At}\alpha_{max}\sin(Bt+\phi_0) \quad (12)$$

With

$$A = \frac{\rho S V d^2}{4 I_y} C_{mq} \quad (13)$$

$$B = 2\pi f = \sqrt{-\frac{\rho S V^2 d}{2 I_y} c_{m\alpha}} \quad (14)$$

$I_y$ is the transverse inertia and $\alpha$ the aerodynamic angle of attack in the plane of incidence. In order to estimate the aerodynamic coefficients, a damped sine wave curve is superimposed on the measurement by means of a curve fitting algorithm. The estimation of the aerodynamic coefficients is performed by identification of the angle of attack evolution model parameters:
1. Curve fitting of the model (12) on the measurement signal
2. Identification of the initial shift $\phi_0$, the period $B$ and the damping factor $A$
3. Calculation of $C_{m\alpha}$ and $C_{mq}$ from equations (13) and (14).

## 3. CAVITY EFFECT INVESTIGATION

A cavity had to be created at the rear part of the model for the Cardan-like joint to be hold by the sting. This artefact may have an impact on the pressure distribution around and inside of the projectile, and thus generate a modification of its attitude with respect to the free flight. To understand and quantify the impact of this cavity, an investigation has to be performed. Due to the complexity of the inner device, the lack of space and no optical access, this study can only be performed in a numerical way.

### 3.1. NUMERICAL INVESTIGATION DESCRIPTION

As illustrated in Fig. 7, in order to predict the influence of the cavity and the holding sting, Reynolds-averaged Navier-Stockes computations were performed at Mach 2 on the full and on the MiRo DREV-ISL projectile. The k-ω SST turbulence model has been employed and the wind tunnel conditions have been taken as boundary conditions, i.e. the pressure $P = 51\,122$ Pa, the Mach number $M = 2$ and the temperature $T = 166.7$ K.

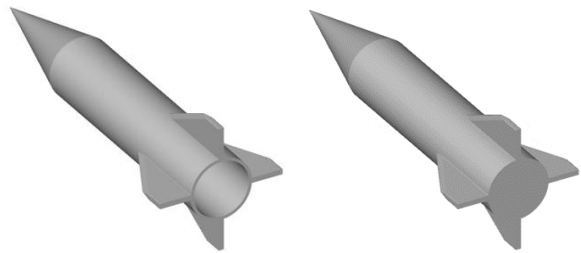

Figure 7. MiRo wind tunnel DREV-ISL (left) and reference DREV-ISL (right)

The simulation were performed with Fluent v19.2 at 3 different angles of attack: 0°, 1° and 2°. For each geometry, a tetrahedral mesh was generated with the ANSYS meshing software and converted to a polyhedral one with the meshing mode of Fluent. Both meshes have the same cell repartitions expecting in the cavity and at the rear of the bases. The number of cells are 6 700 000 and 2 900 000 for the MiRo and reference geometries, respectively (see Fig. 8 and 9).

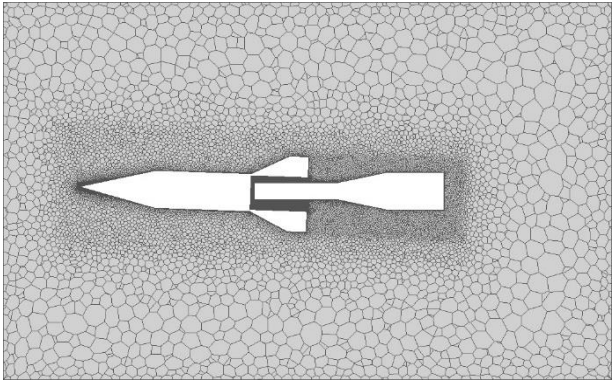

Figure 8. MiRo wind tunnel configuration mesh at 2° angle of attack

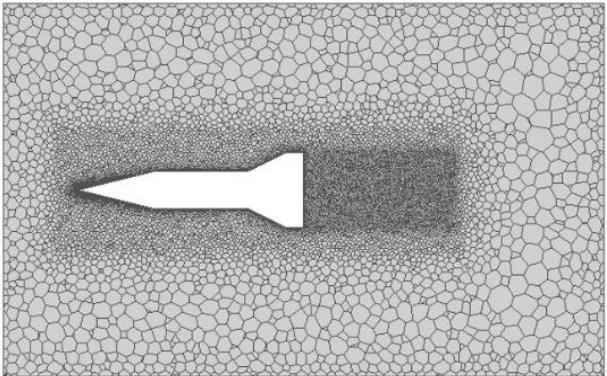

Figure 9. Reference model mesh

## 3.2. PRESSURE DISTRIBUTION ANALYSIS

In this part, the rotation centre around the MiRo holding sting has been set at the centre of gravity, i.e. at a distance of 2.6 calibre from the base. Fig. 10 and 11 give numerical simulation normalized pressure distributions along the projectiles for two angles of attack: 0° and 2°. Those data have been extracted from the wall in the inclination plane that passes through the vertical fins (which is also the plane of symmetry of the model). At 0° angle of attack (Fig. 10), the comparison of the numerical simulation results (blue curve) with already existing wind tunnel pressure measurements [9] (red and black markers) indicates that the forebody pressure prediction show a very good agreement with the experiments. For both geometric configurations (Fig. 7), the respective outer pressure profiles are identical either for the simulations at 0° angle of attack (Fig. 10, blue and orange curves) as for the ones at 2° angle of attack (Fig. 11, both upper and lower pressure profiles). Therefore, it can be considered that the cavity does not impact the outer pressure profile. This property may be due to the supersonic characteristics of the flow, and may not be valid for a transonic or subsonic regime, meaning that additional investigations will have to be done for these lower velocities.

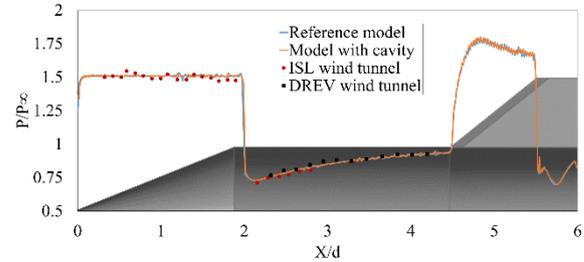

Figure 10. Outer pressure profiles on the projectiles with and without cavity at 0 degree angle of attack

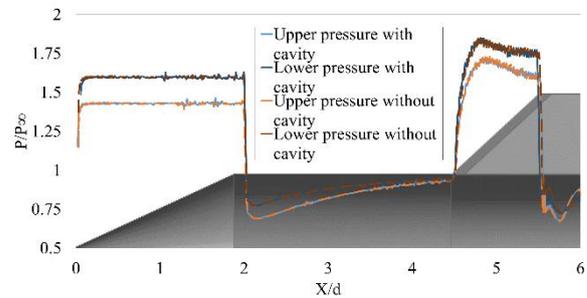

Figure 11. Outer pressure profiles on the projectiles with and without cavity at 2 degrees angle of attack

In the cavity (see Fig. 12), as the symmetry of the simulated case would suggest, the difference between the upper and lower pressure profiles (blue and yellow curves) is close to zero at 0° angle of attack. However, small discrepancies, that can be observed around X/d = 5.5, show that the upper and lower pressure profiles are not perfectly identical, although simulations have shown a good convergence. This observation suggests that the domain close to aperture (domain between 5.5 and 6.0 calibres from the nose) is very turbulent, so that steady state RANS simulations may show some limitations in this specific region. Therefore, for a deeper phenomena investigation, unsteady simulations should be performed, but for the evaluation of the cavity effect on the complete projectile, the order of magnitude of these discrepancies with respect to the outer pressure profiles suggests that the steady state results are acceptable for this investigations.

When the model takes some elevation, a difference between the upper and lower cavity pressure profiles (light blue and red curves of Fig. 12) is obtained, leading to an effect on the attitude of the projectile. However, by calculating the area between the upper and lower pressure curves on the projectile (light and dark blue curves of Fig. 11) and in the cavity (light blue and red curves of Fig. 12), the relative difference of 1% between both integrated pressure profiles suggests that the effect of the cavity on the attitude of the projectile is very low. As this quantity has been calculated from pressure profiles on the inclination plane, for a more precise quantification, additional investigations should be performed first by considering additional parameters (different Mach numbers, different centres of rotations, etc.) and second, by integrating the pressure over the surface.

Furthermore, Fig. 12 shows that once a characteristic depth of almost 1 calibre (X/d = 5) is reached by the flow, the cavity pressure becomes almost constant. This means that the flow velocity becomes close to zero in the deep inner part of the cavity, allowing the pressure to become homogeneous. This flow pattern is completely different from a traditional wake flow, where the pressure profile should effectively be constant (which is not predictable with RANS simulations [13]), but where the total base pressure would be less important, due to a velocity that remains in the recirculation (confirmed by the rear pressure comparison of Fig. 13). This analysis shows that the cavity has a non-negligible effect on the total drag but if the test bench centre of rotation is placed on the projectile's axis, the wind tunnel model attitude will remain very close to the free flight one.

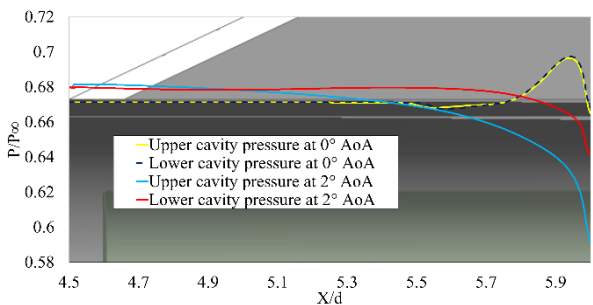

Figure 12. Pressure profiles in the upper and lower cavity at 2 and 0 degrees angle of attack

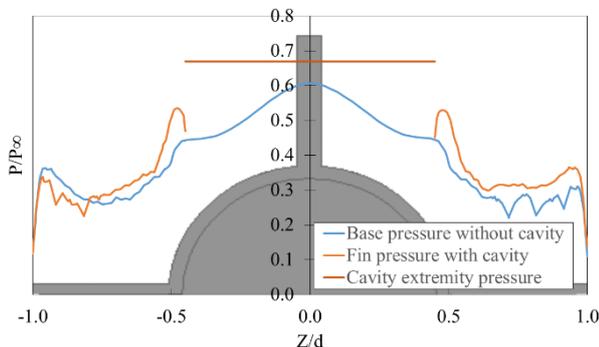

Figure 13. Base pressure profile at 0 degree angle of attack

All these conclusions can be validated by comparing the global aerodynamic coefficients from Tab.1. The normal force coefficient slope $C_{N\alpha}$ and the pitch moment coefficient derivative $C_{m\alpha}$ that are obtained by CFD with and without cavity are almost identical and also very close to the wind tunnel measurements [9]. Only the drag force coefficient offset shows a value that decreases from 10% when the cavity and the holding sting are added. Complementary cavity effect investigations are still to be done on a complete projectile flight domain, especially in the subsonic regime.

|  | CFD with cavity | CFD without cavity | Wind tunnel [9] |
|---|---|---|---|
| $C_{X0}$ | 0.445 | 0.511 | 0.53 |
| $C_{N\alpha}$ | 6.95 | 7.06 | 7.05 |
| $C_{m\alpha,base}$ | 14.12 | 14.24 | 14.61 |

Table 1. Global aerodynamic coefficients comparison with and without cavity

## 4. WIND TUNNEL MIRO EXPERIMENTAL CAMPAING

### 4.1. WIND TUNNEL TEST CONDITIONS

The experiments were performed in ISL's trisonic blow down wind tunnel [12] (See Fig. 14), which allows to perform investigations in a Mach range from 0.5 to 4.5. The tests presented in this paper were performed at Mach 2 with a stagnation pressure of 4.1 bars and a total temperature of 298°K.

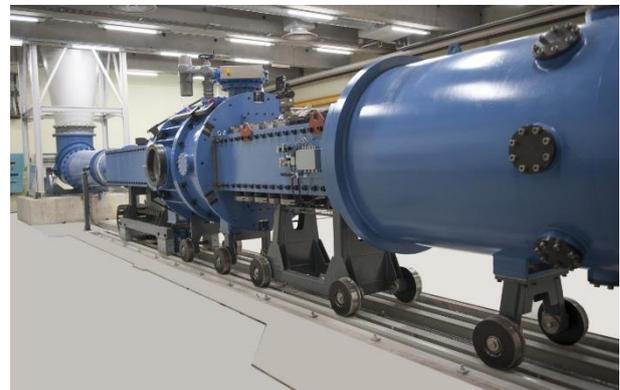

Figure 14. ISL's trisonic wind tunnel

### 4.2. OPTICAL SETUP

The attitude of the projectiles was recorded with 2 Photron SA-Z high-speed cameras. They are able to record 20000 frames per second in a full frame format (1024x1024 pixels) with an exposure time of 0.5μs. In order to avoid the motion blur, which would decrease the detection precision during the post-treatment, a very low exposure time is necessary to ensure a clear image regardless of the model's attitude. For this reason, four Dedotech Dedolight 400D DLH400D professional cool lights have been used to illuminate the black model on which white dot markers have been put. Both camera lenses have a focal length of 105 mm. The cameras were placed at a distance of 1.2 meters of the model and spaced of 0.6 meter. In this configuration, as illustrated in Fig. 15, the angle between both optical axes is 30 degrees.

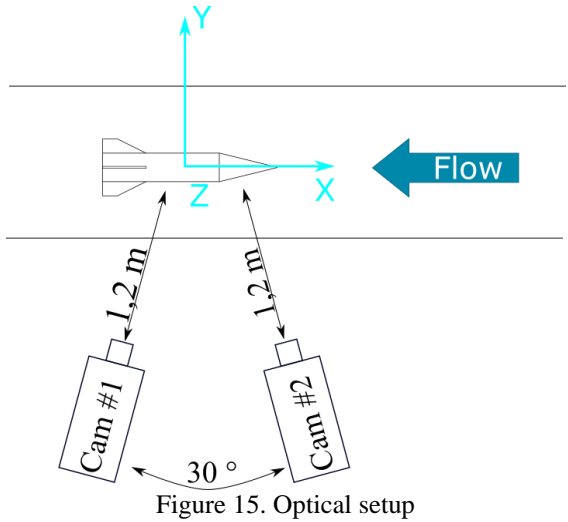

Figure 15. Optical setup

### 4.3. EXAMPLE OF NON-PERTUBED FLIGHT

In order to check the stability of the DREV-ISL at Mach 2 and to test the mechanical behaviour of the test bench in supersonic flow, a first test has been performed without mechanical perturbation. As illustrated in Fig. 16, the stable centre of gravity position ($10^{-4}$ m displacement amplitude on the 3 axis) during the blow down shows that the mechanical test bench holds the stress in the test conditions. The noise on the projectile's attitude on Fig. 18 is a combination of algorithmic and mechanical noises. As awaited, the green and red curves on Fig. 17 indicate that there are almost no pitch and yaw motions (maximum angular rotation is order of ±0.3 degrees) but a random roll motion is obtained due to aerodynamic perturbations. This last observation indicates, that the friction of the bearings is low enough to allow the model to spin gently.

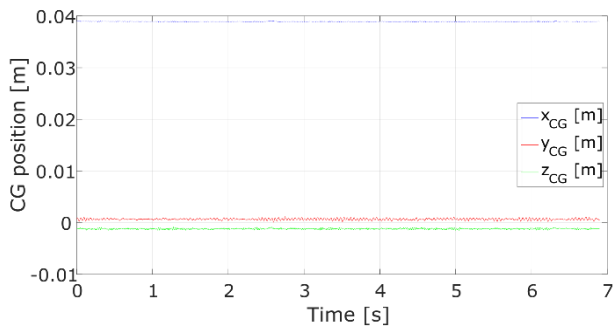

Figure 16. Centre of gravity position without mechanical perturbation (test #1)

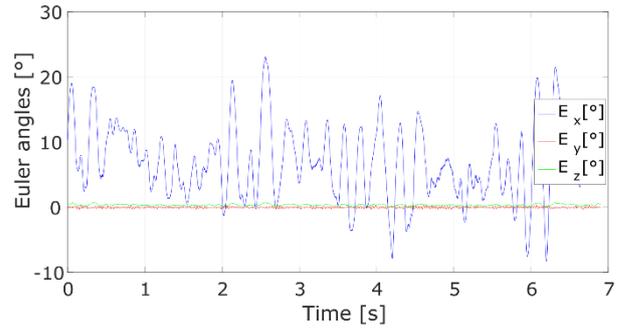

Figure 17. Euler angles without perturbation (test #1)

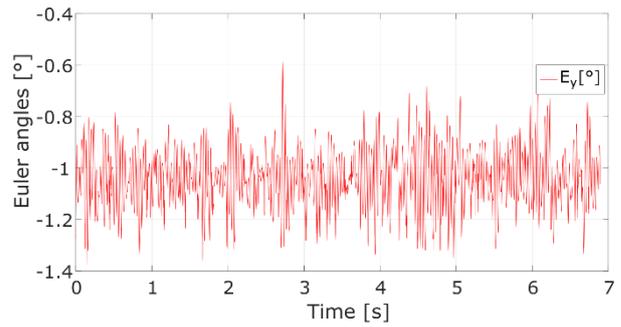

Figure 18. Pitching motion without perturbation (test #1)

### 4.4. AERODYNAMIC COEFFICIENT DETERMINATION

The method described in part 2.3 was applied on this experiment to determine the $C_{m\alpha}$ and $C_{mq}$ aerodynamic coefficients. For repeatability investigation, three experiments were performed. In order to avoid a rotating unbalance the MiRo test bench was set up so that the Cardan joint centre of rotation was superimposed on the centre of gravity of the model. For this reason, the attitude submitted by the model is only due to the aerodynamic loads. Like shown on Fig. 19, high amplitude damped oscillations are observed. The polar diagram on Fig. 20 also shows that the model describes an almost planar motion.

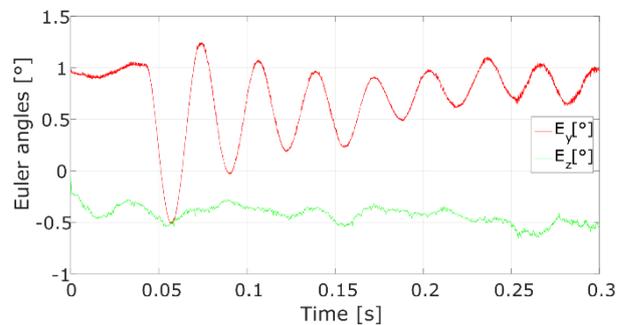

Figure 19. Euler angles obtained for the perturbed experiment.

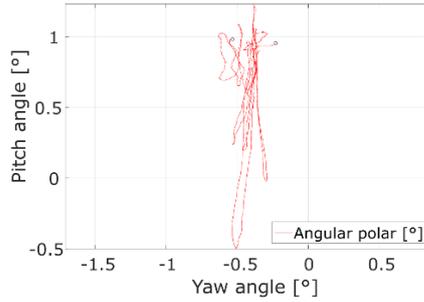

Figure 20. Angular polar curve obtained for the perturbed experiment

As soon as the attitude of the projectile is planar, the pitching moment derivative coefficient $C_{m\alpha}$ estimation is based on a Fourier transform performed on the pitch angle signal obtained in Fig. 19. This procedure allows to estimate the signal frequency $f$, and to estimate the value of $B$ (Eq. 14), which finally allows to calculate the $C_{m\alpha}$.

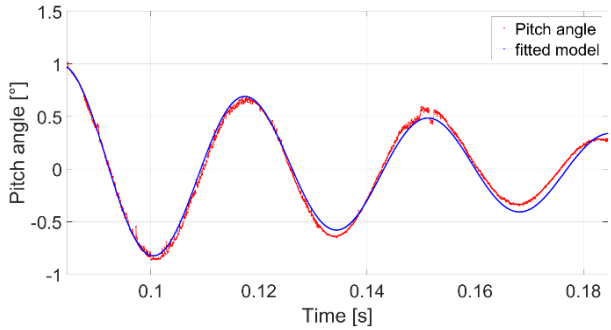

Figure 21. Pitch angle measurement and model curve fitting

Fig. 21 shows the pitch angle measurement and its respective fit with the curve of the model given in Eq. 12. The first few milliseconds, are ignored so as not to take the initial movement produced by the perturbation device into account. For this example, the damping factor $A$ (Eq. 13) is equal to 185 rad.s$^{-1}$ and the angular frequency $f$ to 29.6 Hz.

|  | Frequency (Hz) | $C_{m\alpha}$ MiRo | $C_{m\alpha}$ Free-flight [1] | $C_{mq}$ MiRo | $C_{mq}$ Free-flight [1] |
| --- | --- | --- | --- | --- | --- |
| Test #1 | 29.6 | -4.52 |  | - |  |
| Test #2 | 29.86 | -4.55 | -4.29 | -37.1 | -40 |
| Test #3 | 29.53 | -4.45 |  | -37.3 |  |

Table 2. Aerodynamic coefficients measurement results

Tab. 2 shows that the experiment presents a very good repeatability and both static and dynamic pitch moment coefficients are very close to the free flight identification. Moreover, it is very important to notice that the $C_{mq}$ is a coefficient that is quite complicated to be measured. A relative difference of only 7% for the results presented in Tab. 2 is very promising for the MiRo setup development to be continued.

### 4.5. MARKERS EVOLUTION

For the first wind tunnel campaigns (tests #1 and #2), manually set dots have been employed as markers, and Gauss fittings were employed to detect their respective centre position. However, for a same dot, if its shape is not perfect, small displacements of the identified centre location can occur between consecutive image pairs. As shown on Fig. 22, this artefact affects the results by generating algorithmic noise.

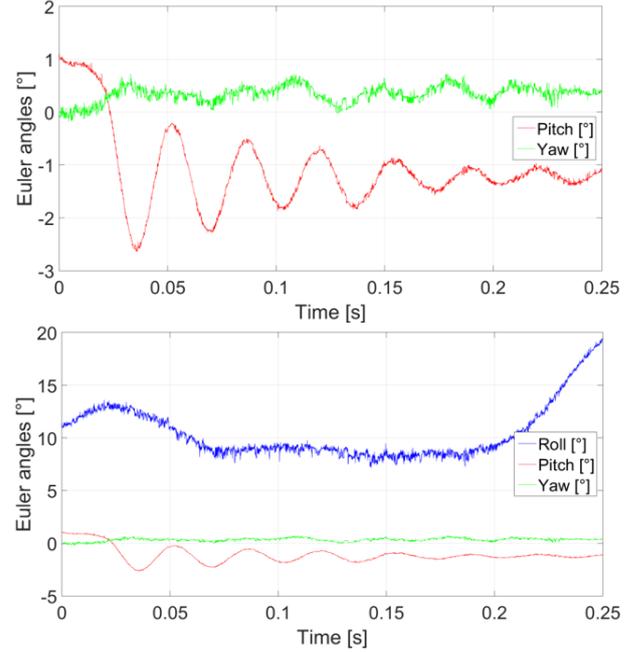

Figure 22. Euler angles measurements with dot detections

In the third test campaign (results given for test #3), dots were replaced by Secchi markers and the Gauss fitting algorithm by Harris and Stephen's corner detection method [7]. This modification improved the centre detection and, in the same time, notably reduced the noise on the pitch and yaw signals like shown on Fig 24. Furthermore, by comparing Fig. 16 with Fig. 23, the noise amplitude on the centre of gravity position signal has been evaluated and has decreased by a factor of 10$^{-2}$. Only the noise level on the roll signal has not decreased. This observation can be explained by the thickness of the fin to be too low for the mesh to 3D markers scatter points fitting algorithm to perform a precise fit. In fact, when the algorithm tries to fit the mesh on the markers 3D points, if some calculated marker coordinates are not precise enough, the mesh projection can jump from one fin face to the other, generating a parasitized roll signal. However, a traditional filtering process (not done for all the results presented in this article) could be a simple solution if this remaining noise would generate difficulties for data exploitation.

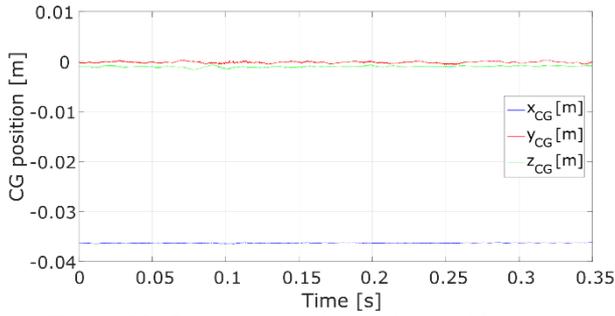

Figure 23. Centre of gravity position with corner detections

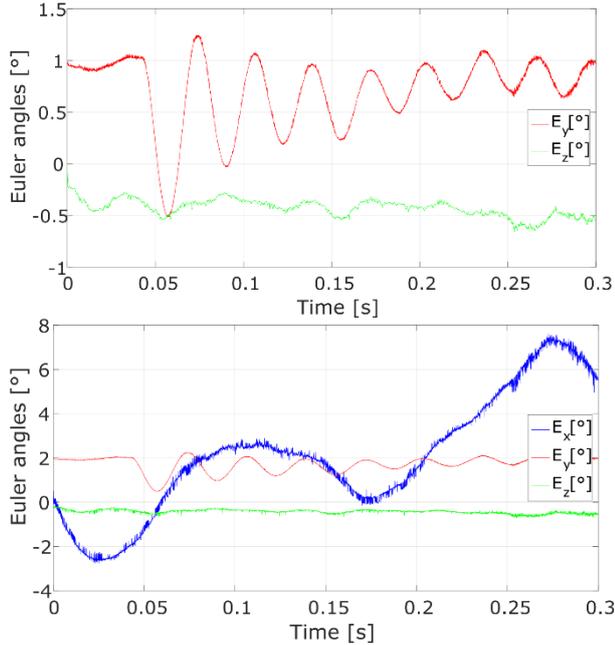

Figure 24. Euler angles measurements with corner detections

## CONCLUSION AND OUTLOOK

A new wind tunnel concept validation experimental setup called MiRo has been developed. This method is based on the stereovision technique that is applied on a rotation free mechanical test bench mounted in the wind tunnel. This device showed its efficiency for the determination of dynamic and static aerodynamic coefficients on a reference rocket model, called the DREV-ISL. A first evaluation of the impact of the holding sting cavity on the model's attitude in a supersonic flow has been performed. This investigation has shown that the cavity has a quasi-negligible effect on the angular motion, and therefore, on the measurements of the pitch moment coefficient slope ($C_{m\alpha}$) and the pitch damping coefficient ($C_{mq}$) as long as the Cardan joint is precisely centred on the model's axis. Other investigations should be performed, first with Unsteady Reynolds Averaged Navier-Stockes (URANS) simulations to get a better understanding of the physical flow phenomena that occur in the cavity and, secondly, for other Mach numbers, rotation centre locations and geometries, to confirm these conclusion in a wider range of flight points. The $C_{m\alpha}$ and $C_{mq}$ measurements obtained by the MiRo wind tunnel test campaign on the DREV-ISL is in good agreement with already existing free-flight measurements. These results show that this new methodology is promising and that the development can move to its second phase, namely the one which deals with the uncoupled rear part of the tested model.